\documentclass[12pt]{article}
\newcommand{\beq}{\begin{equation}}
\newcommand{\eeq}{\end{equation}}
\newcommand{\beqs}{\begin{eqnarray}}
\newcommand{\eeqs}{\end{eqnarray}}
\newcommand{\laa}{\lambda_{IIA}}
\newcommand{\lop}{\lambda_{I'}}
\newcommand{\lhet}{\lambda_{E_8}}

\newcommand{\rhet}{R_{E_8}}
\newcommand{\da}{{\dot a}}
\newcommand{\db}{{\dot b}}
\begin{document}

\begin{titlepage}
\begin{flushleft}
       \hfill                      {\tt hep-th/9610082}\\
       \hfill                       UUITP-24/96\\
       \hfill                       October 1996\\
\end{flushleft}
\vspace*{3mm}
\begin{center}
{\LARGE The Heterotic Life of the D-particle.\\ }
\vspace*{12mm}
{\large Ulf H. Danielsson\footnote{E-mail: ulf@teorfys.uu.se} \\
\vspace{4mm}
Gabriele Ferretti\footnote{E-mail: ferretti@teorfys.uu.se} \\
\vspace{4mm}
{\em Institutionen f\"{o}r teoretisk fysik \\
Box 803\\
S-751 08  Uppsala \\
Sweden \/}}\\
\vspace*{10mm}
\end{center}

\begin{abstract}
We study the dynamics of D-particles (D0-branes)
in type I' string theory and
of the corresponding states in the dual heterotic description.
We account for the presence of the two 8-orientifolds
(8 dimensional orientifold planes) and
sixteen D8-branes by deriving the appropriate quantum mechanical system.
We recover the familiar condition of eight D8-branes for
each 8-orientifold.
We investigate bound states and
compute the phase shifts for the scattering of such states
and find that they agree with the expectations from the
supergravity action.
In the type I' regime we study the motion transverse to the
8-orientifold and find an interesting cancellation effect.
\end{abstract}
\end{titlepage}

\section{Introduction}

String solitons carrying RR charges, known as D-branes, \cite{P1, PCJ}
have proven to be of the utmost importance in testing and making use
of the various string dualities that have emerged in the last two years
\cite{HT, T, W1, PW, S6, HW2, W2, T2, HW1}\footnote{An early review of string
solitons is \cite{DKL}, some recent reviews of various
aspects of duality are \cite{P2, S7, D2, D3}.
Dirichlet boundary conditions
for open strings have been studied, for instance, in \cite{DLP, L1, G}.
Among the many applications of D-brane techniques, we should also
mention black hole entropy \cite{SV, DM, CM, HS2, GKP, M} and
string/D-brane interaction (e.g. gravitational lensing) \cite{KT, GHKM,
HK, H2}.}. More recently, these D-branes have also provided
probes \cite{us, KP, DL, D1, BDS, DKPS, S1} of the geometry of spacetime.

The simplest of all D-branes is obviously the D-particle (D0-brane)
in type IIA string theory. From the ten dimensional
point of view it corresponds to a
charged, extremal black hole, while from the eleven dimensional
M-theory point of
view it is a Kaluza-Klein state where the quantized charge is related to
the quantization of momentum in the compact direction. The D-particle is
just an eleven dimensional graviton traveling around a circle.

Clearly such a system deserves a careful study. Perhaps important lessons
can be learnt about the elusive M-theory? Such a study has been initiated
in \cite{us,KP,DKPS}. There it was found that not only the mass and charge
of the D-particle has an eleven dimensional interpretation but also the
dynamics is best understood in these terms. The characteristic length and
energy scales are both eleven dimensional. This suggests the fascinating
possibility that D-particles might be used to {\it formulate} M-theory,
\cite{BFSS}. A necessary condition for this idea to be applicable
is Hull's analysis \cite{H1}
suggesting that these objects dominate in the strong coupling limit of
the type IIA string.

In this paper we continue the study in the case of configurations relevant to
the type I' and heterotic theories.
To do this we construct the type I' theory
as a type IIA orientifold \cite{S3, H3, PS, GP},
where the last space coordinate is a
$S^1/ Z_2$ orbifold. The type I' D-particle is then the T-dual of the
type I D-string. We study how such
D-particles interact in this environment. By using type I/heterotic duality
\cite{PW, HW2, HW1} we can identify these states with the Kaluza-Klein modes of
the
heterotic string around a circle and investigate some of their
properties as well.

The outline and the main results of the paper are as follows.
In the next section, we begin with
a review of the type IIA D-particle. We point out some of the most
interesting results obtained in previous work, in particular, the
importance of the eleven dimensional Planck scale. In section three,
we derive the quantum mechanical system that describes the
interactions of the type I' D-particles.
In section four we discuss the physical interpretation of this model;
we identify the flat directions and argue about the existence of
non BPS resonances. In section five we present the results
in terms of effective potentials. We study the scattering of such states
against the 8-orientifold and against each other. The potential is
$O(v^2)$ and $O(v^4)$ respectively.
In section six we try to extrapolate the results to the heterotic
string using string dualities. We see that the dynamics is
reproduced by the supergravity action.
There, we also find that the eleven dimensional
Planck scale plays an important role in the
type I' theory and speculate that the same might happen in
the heterotic limit.

\section{Review of the type IIA D-particle}

In this section we briefly review some of the results of \cite{us,KP,DKPS}
concerning the type IIA D-particle. Consider a collection of $N$ such
D-particles;
their low energy, short distance, effective action
\cite{W3} is given by the quantum
mechanical system obtained by the dimensional reduction
\cite{BSS} of a (9+1)-dimensional
$U(N)$ Super Yang-Mills theory all the way down to 0+1 dimension. The
$U(1)$ subgroup factors out and describes
the free motion of the center of mass,
leaving a non trivial $SU(N)$ dynamics. The reduced system has 16
supercharges\footnote{The notation is as in \cite{us} and will not be repeated;
$E_i^a$ is the canonical conjugate of $A_i^a$,
$\psi^a$ are 16 component spinors satisfying
``gamma matrix'' type anticommutation relations and $f^{abc}$are the structure
constants of $SU(N)$.}
\beq
       Q_\alpha = \sqrt{\laa}
       \gamma_{i\alpha\beta}\psi^a_\beta E_i^a - {1\over {2\sqrt{\laa}}}
       f^{abc} \gamma_{ij\alpha\beta} \psi^a_\beta
       A_i^b A_j^c, \label{IIAsupercharges}
\eeq
anticommuting to give the hamiltonian
\beq
        H =  \frac{\laa}{2} E_i^{a2} - {1\over 2}  i f^{abc}
             A_i^a \psi^b\gamma_i\psi^c
          + {1\over {4\laa}} \left(f^{abc} A_i^b A_j^c\right)^2
                    \label{IIAhamiltonian}
\eeq
up to a gauge transformation.

The hamiltonian (\ref{IIAhamiltonian}) should have one
marginally stable bound state
at threshold as required by the relation between type
IIA and M-theory \cite{T,W1, W2}.
While by now nobody doubts this fact, mostly due to the
arguments of Sen \cite{S4, S5},
a ``direct'' proof of this fact is still missing.
What \emph{can} be done is to show the
existence of resonances whose energy depends
on the string coupling as $E \approx (\laa)^{1/3}$ in string units (i.e.
$\alpha'_{IIA}=1$). As the binding is due to open
strings stretching between pairs of
D-particles and each string contributes with a
potential that grows linearly with the
separation, the average size of such resonances is $r\approx (\laa)^{1/3}$.
Also, as the mass of each constituent is $1/\laa$,
the orbital motion is non relativistic at weak
coupling, with typical orbital velocity $v\approx \laa^{2/3}$.

This is consistent with the picture emerging from the one loop string
calculation of \cite{B} (see also \cite{BS, FPR, L2, BCFPR}).
There, one studies the scattering
of two D-branes and sees the growth of
the absorptive part $\mathrm{Im}(\delta)$ of the phase shift
as the impact parameter $r$ becomes shorter than
the square root of the velocity\footnote{This region is referred
to as ``the stadium'' in \cite{DKPS}.} (always in string units).
This is precisely the same scaling found for
the above resonances, $r\approx \sqrt{v} \approx \laa^{1/3}$,
confirming the idea that these are the states responsible
for the inelasticity.

An even more intriguing outcome of the calculation is the appearance of the
11~dim Planck length $l_{11}\approx\laa^{1/3}$ as a dynamical length.
A hint that the D-particle ``remembers'' its 11~dimensional
origin can already be seen by expressing
the resonance energies in units appropriate to the 11~dimensional theory as in
\cite{us}, but definitive evidence was presented in \cite{KP,DKPS} by pointing
out that there is a range of velocities
where the D-particles have non relativistic
dynamics and energies below the string mass while having momentum of order
$1/l_{11}$. At weak coupling, $l_{11}$ is shorter than the string length,
indicating that D-particles (actually, D-branes in general)
may be used to probe sub-stringy
distances in the spirit of an earlier proposal of Shenker
\footnote{However, the
11~dim Planck length is still larger than Shenker's original proposal for a
minimal length that was $l_{min}\approx \laa$.
This length coincides with the 4~dim Planck
length and, amusingly, with the M-theory compactification radius $R_{11}$ in
string units.}\cite{S2}.

In \cite{DKPS} the scattering of two D-particles was also studied in the
eikonal (i.e. background gauge) approximation,
also showing agreement with the calculation of
Bachas \cite{B}. The main physical results were that the moduli space for such
objects remains flat\footnote{I.e., the potential between two type IIA
D-particles is $O(v^4)$ due to the large number of supersymmetries.}
and that $l_{11}$ also governs the scattering amplitude via
\beq
     f(k,\theta) \approx e^{-\sqrt{2}\sin(\theta/2)(k l_{11})^{3/2}}.
\eeq
We shall see the counterpart of all these statements in the type I'
and Heterotic case.

\section{The type I' D-particle}

In this section we work out the lagrangian
and the hamiltonian descriptions for
the type I' D-particle to be used for dynamical studies in the following
sections. We
begin with identifying the appropriate degrees of freedom by a straightforward
application of the formalism developed in \cite{GP}. The type I' theory is an
orientifold quotient of type IIA theory obtained by T-dualizing the usual
construction of type I string theory as an orientifold of type IIB:
\beq
     I = \frac{IIB}{\{1,\Omega\}} \stackrel{T}{\longrightarrow}
     I' = \frac{IIA}{\{1, \Omega R\}}.
\eeq
In both cases the orientifold group is isomorphic to $Z_2$, where the non
trivial element is simply world-sheet parity $\Omega$ in the type IIB case
and the composition of world-sheet parity with the reflection $R$ along,
say, $X^9$ in the type IIA case. The direction $X^9$ is assumed to be
compactified on a circle and it
is on this circle that T-duality acts as usual by inverting the radius.
The two fixed points of the action of the reflection $R$ on the circle are
identified as the positions of the 8-orientifold.
Type I D-strings wrapping around the circle are T-dualized to the
D-particles we wish to consider.
By symmetry, these D-particles are originally on the
8-orientifold but we will also consider their dynamics in the perpendicular
direction $X^9$ by allowing for Wilson lines.

In addition to D-particles and orientifolds, the familiar consistency
conditions on the
RR charges require the presence of 16 D8-branes, the T-dual of the type I
D9-branes. We consider only the simple situation of 8 such D8-branes
sitting near each of the two 8-orientifolds, so that there
is no net RR background charge
in the bulk. We can introduce Wilson lines for these D8-branes as well,
corresponding to small displacements away from the 8-orientifold.
This requires the addition of fermionic bilinears
(``bare masses'') to the action for the  D-particle
coming from the quantization of open strings in the 0-8 sector.
This is important for a full cancellation of the vacuum energy of
the fast modes.

Consider having $N$ D-particles. The Chan-Paton indices $I,J$ will run from
$1$ to $2N$ because of the presence of the mirror images.
In the 0-0 open string massless sector,
the action of $\Omega R$ is given by \cite{GP}:
\beqs
     b_{-1/2}^{0,9}|IJ> &\to& - M_{II'}b_{-1/2}^{0,9}|J'I'>M_{J'J}^{-1}
          \nonumber\\
    \Omega R:\quad  b_{-1/2}^{1,2\cdots 8}|IJ> &\to&
         M_{II'}b_{-1/2}^{1,2\cdots 8}|J'I'>M_{J'J}^{-1}\nonumber\\
     |\alpha IJ> &\to& \gamma^9_{\alpha\beta}
        M_{II'}|\beta J'I'>M_{J'J}^{-1}.\label{consistency}
\eeqs
A few words of explanation are in order. 

The bosonic coordinates are obtained by
acting with the operators $b_{-1/2}^\mu$ on the NS vacuum $|IJ>$ and the
fermionic coordinates are given by the GSO projected R vacuum
$|\alpha IJ>$, where $\alpha=(a, \da)\in \mathbf{8}_s + \mathbf{8}_c$.
The matrix $M$ is a $2N\times 2N$ symmetric and
unitary matrix whose only non zero
entries are those connecting a D-particle
with its mirror and the $16\times 16$ matrix $\gamma_9$ is related to the
$32\times 32$ Dirac matrices of $\mathrm{spin}(9,1)$ as
$-\Gamma^0\Gamma^9 = \gamma_9\otimes\sigma_3$ with the conventions of
\cite{us}. Later we will also
need the explicit expression of the matrices $\gamma_1, \cdots, \gamma_9$ in
terms of their own $8 \times 8$ ``Pauli'' matrices:
\beq
    \gamma_i = \pmatrix{0 & \sigma^i_{a\da}\cr \sigma^i_{\db b} & 0 \cr}
    \quad i=1,\cdots, 8\qquad \gamma_9=\pmatrix{\delta_{a b} & 0\cr
                   0 & -\delta_{\da \db} }.
\eeq

Armed with those tools we can easily work out the relevant degrees of freedom
by
requiring that we keep only states invariant under $\Omega R$. The surviving
bosonic coordinates are
\beq
        A_{0,9}^{IJ} \quad X_{1,\cdots, 8}^{IJ}\quad\mathrm{and}\quad
        x_{1,\cdots, 8},
\eeq
where $A_0$ and $A_9$ are antisymmetric in the indices $I$ and $J$, giving  the
gauge group $SO(2N)$, whereas the remaining coordinates split into a traceless
symmetric representation $X_i^{IJ}$ and a singlet $x_i$. Similarly, the
surviving fermionic coordinates are
\beq
        S_a^{IJ} \quad S_{\da}^{IJ}\quad\mathrm{and}\quad  s_{\da},
\eeq
where $S_a^{IJ}$ is in the adjoint (i.e., antisymmetric), and
$S_{\da}^{IJ}$ is in the
traceless symmetric and $s_{\da}$ is again a singlet. The bosonic and fermionic
singlets describe the collective motion of the centre of mass
parallel to the 8-orientifold and we will
ignore them in what follows. We can represent the action of $SO(2N)$
on the various fields by
simple commutators or anticommutators; e.g. $[A_9,X_i]$ is itself traceless and
symmetric, and so is $\{S_a, S_\da\}$.

The action, the  hamiltonian and the 8 supercharges
are fixed uniquely by group theory
and supersymmetry. It is straightforward to obtain them by generalizing the
formulas of \cite{us, KP, DKPS}. The action is:
\beqs
        &&S=\int dt\;\mathrm{Tr}\Bigg\{\frac{1}{\lop}
        \bigg( \frac{1}{2} \dot X_i^2  -\frac{1}{2} \dot A_9^2
        - \dot A_9 [A_0, A_9] + \dot X_i [A_0, X_i] -
        \frac{1}{2}[A_0, A_9]^2 \nonumber \\
       &&+ \frac{1}{2}[A_0, X_i]^2 - \frac{1}{2}[A_9, X_i]^2 +
        \frac{1}{4}[X_i, X_j]^2 \bigg)  +
        \frac{i}{2} \bigg( S_\da \dot S_\da  - S_a \dot S_a -
        S_a[A_0, S_a]\nonumber \\
        &&+ S_\da[A_0, S_\da] + S_a[A_9, S_a] + S_\da[A_9, S_\da] -
        2X_i \sigma^i_{a\da} \{S_a, S_\da\} \bigg) \Bigg\}. \label{action}
\eeqs
{}From (\ref{action}), the hamiltonian and the Gauss' law can be derived. We
present only the hamiltonian in the temporal gauge $A_0 =0$.
Denoting by  $E_9$ and $P_i$
the momenta canonically conjugated to $A_9$ and $X_i$ we have:
\beqs
        H &=& \mathrm{Tr}\Bigg\{ \lop \bigg(
        \frac{1}{2} \dot P_i^2  - \frac{1}{2} \dot E_9^2  \bigg) +
        \frac{1}{\lop}\bigg( \frac{1}{2}[A_9, X_i]^2 -
        \frac{1}{4}[X_i, X_j]^2 \bigg)  \nonumber \\&& +
        \frac{i}{2} \bigg( -S_a[A_9, S_a]  - S_\da[A_9, S_\da] +
        2X_i \sigma^i_{a\da} \{S_a, S_\da\} \bigg) \Bigg\}. \label{hamiltonian}
\eeqs
Finally we can write down the 8 (and not 16 as in the type IIA case)
supercharges
whose anticommutation relations give (\ref{hamiltonian}) up to a gauge
transformation:
\beq
       Q_a =\mathrm{Tr}\Bigg\{ \sqrt{\lop}\bigg( \sigma^i_{a\dot a}S_\da P_i
       - S_aE_9\bigg) + \frac{1}{2\sqrt{\lop}} \bigg( \sigma^{ij}_{ab}
       X_i[S_b, X_j] + \sigma^i_{a\da}S_\da[A_9, X_i] \bigg) \Bigg\}.
       \label{supercharges}
\eeq

Equations (\ref{action},\ref{hamiltonian},\ref{supercharges}) define
the system we wish to
study in the following sections, for general $N$ and for
$N=2$ (gauge group $SO(4)$) in particular.

\section{Bound states and flat directions}

The moduli space of the system is defined by the gauge inequivalent
solutions to
\beq
         [A_9, X_i] = [X_i, X_j] =0.
\eeq
The flat directions associated with $A_9$ play a special role in that they
define the position of the D-particles relative to the 8-orientifold
or, equivalently, relative to
their mirror images. To fix the ideas, consider the case of  two D-particles.
The two elements of the Cartan subalgebra of $SO(4)$ can be written as
\beq
        H^1 = \frac{1}{2}\left (\matrix{
                    0 & 1 & 0 & 0 \cr
                   -1 & 0 & 0 & 0 \cr
                    0 & 0 & 0 & 0 \cr
                    0 & 0 & 0 & 0 \cr
                    }\right ) \quad\mathrm{and}\quad
        H^2 = \frac{1}{2}\left (\matrix{
                    0 & 0 & 0 & 0 \cr
                    0 & 0 & 0 & 0 \cr
                    0 & 0 & 0 & 1 \cr
                    0 & 0 & -1& 0 \cr
                    }\right ) .
\eeq
The configuration $A_9 = a_9^1H^1 + a_9^2H^2$ represents
two D-particles whose
distances from the 8-orientifold are $a_9^1$ and $a_9^2$ respectively. In the
generic situation, the only traceless symmetric matrix that commutes
with $A_9$ above is
\beq
        M^0 = \frac{1}{2\sqrt{2}}\left (\matrix{
                    1 & 0 & 0 & 0 \cr
                    0 & 1 & 0 & 0 \cr
                    0 & 0 &-1 & 0 \cr
                    0 & 0 & 0 &-1 \cr
                    }\right ),   \label{mzero}
\eeq
and the coordinates $x_i$ in $X_i = x_i M^0$ represent the relative distance of
the D-particles in the eight directions parallel to the 8-orientifold. These
coordinates, together with those of the centre of mass, fix the
configuration of the system.

In general for $SO(2N)$ there will be $N-1$ neutral directions (corresponding
to a weight $(0,...,0)$ in Dynkin basis) that give rise to flat directions.
This is precisely what is needed to account for the relative positions
of the $N$ D-particles parallel to the 8-orientifold.
It is a familiar fact that, for particular ``degenerate'' points in the Cartan
subalgebra, more general configurations for the $X_i$'s are possible.
That is, new flat directions open up that reduce
the rank of the gauge group. In terms of weights we
must then turn on the $X_i$ in directions that are
not neutral, see e.g \cite{Par}.
Semiclassically, this corresponds to ``locking'' the system in a specific
configuration along the ninth coordinate, but it is not clear whether
it is possible to prepare the system in such configurations.
An interesting example of this phenomenon is the configuration
$X_i = x_i \tilde M$ that for
\beq
    \tilde M = \frac{1}{\sqrt{2(a^2 + b^2 + c^2 + d^2)}}
    \left (\matrix{
                    a & 0 & 0 & 0 \cr
                    0 & b & 0 & 0 \cr
                    0 & 0 & c & 0 \cr
                    0 & 0 & 0 & d \cr
                    }\right ),\quad a+b+c+d=0\;\hbox{~~all different}
\eeq
breaks the gauge group entirely, therefore locking the two D-particles on the
8-orientifold.

The existence of resonances above BPS saturation can be investigated with
exactly the same tools as in \cite{us} to which we refer for all
the details. What one does is to
fix the temporal gauge $A_0=0$ and to separate the  $18 N^2 + 7N -8$ bosonic
directions\footnote{$\mathrm{dim(Adj}(SO(2N))\mathrm{)} + 8\;
\mathrm{dim(Sym}(SO(2N))\mathrm{)} = 18 N^2 + 7N -8 $.} into $2N^2 + 7N  - 8$
slow modes\footnote{The $N + 8(N-1)$ dimensional moduli space plus
$\mathrm{dim(}SO(2N)\mathrm{)} - \mathrm{rank(}(SO(2N)\mathrm{)}$ gauge
directions.} and the remaining $16 N^2$ fast modes, i.e., those modes for which
the hessian of the bosonic potential is (generically) non degenerate.

The ground state energy for the bosonic fast modes is cancelled by the ground
state energy of the fermions, as expected by supersymmetry, leaving only a
confining linear potential independent of the string coupling. This is
interpreted as coming from on shell strings stretching between the two
D-particles.
To be precise, this cancellation is not exact unless we also
add the contribution from the D8-branes as we will show in the
next section.

When the linear potential is included in the effective action for the slow
modes, as prescribed by the Born-Oppenheimer approximation,
one sees the existence of
bound states above threshold whose dependence on the string coupling can be
computed exactly by the same scaling argument as in \cite{us}.
The following table
summarizes the scaling of the mass $M$, the binding energy $E$, the orbital
velocity $v$ and the orbital momentum $p$ in the units appropriate to the
relevant string theories and ``M-theory'' in the notation of Horava and Witten
\cite{HW1}; notice the simple dependence of all the quantities in the M-theory
units.

For later use, we also include the heterotic string units in the table.
We will come back to this in section 6.

\vskip1cm
\begin{center}
\begin{tabular}{c|c|c|c|}
        & $I'$ & $E_8\times E_8$ & M-theory \\
    &&&\\ \hline &&&\\
    $M$ & $1/\lop$          & $1/ \rhet$              & $1/R_{11} $ \\
    &&&\\ \hline &&&\\
    $E$ & $\lop^{1/3} $    & $\rhet/\lhet^{2/3}$ & $R_{11}$ \\
    &&&\\ \hline &&&\\
    $v$ & $\lop^{2/3}$    & $\rhet/\lhet^{1/3}$ & $R_{11} $ \\
    &&&\\ \hline &&&\\
    $p$ & $1/ \lop^{1/3}$ & $1/ \lhet^{1/3}$      & $1$ \\
    &&&\\ \hline
\end{tabular}
\end{center}
\vskip1cm

\section{Scattering and the metric on the moduli space}

We now move on to the related issue of the scattering of these objects.
We use the same techniques as in \cite{KP, DKPS} i.e., the computation
of the potential and the quantum mechanical phase shifts in the eikonal
approximation. There are two simple situations we wish to consider:
the first is the scattering of one D-particle off ``the
plane''\footnote{We refer to the composite system of 8-orientifold plus 8
D8-branes and their mirror images simply as ``the plane'' for
conciseness.}, the second is the
scattering of two D-particles against each other and parallel to the plane.
In the first case, we will find a potential
$\mathcal{V}\approx v^2 r^{-3}$, where $v$ is
the velocity of the D-particle towards the plane and $r$ the distance
between the two objects. The fact that the potential vanishes at zero
velocity is due to the 8 supersymmetries of the system and to the
cancellation between the contributions form the 8-orientifold and
the D8-branes. In the second case, we find that the potential is of
order $v^4 r^{-7}$ exactly as in the type IIA case.

In the case of a D-particle moving  towards the plane, the relevant
gauge group is $SO(2)$, whose unique generator we represent by
$(i/2) \sigma^2$. The singlets described in section 3 decouple as usual
but we are still left with two
more bosonic degrees of freedom for each parallel direction: $X_i = 1/2 (x_i
\sigma^1 + \tilde x_i \sigma^3$). The fermions are subjected to the
same treatment as the bosons, with eight real Grassmann degrees of freedom
from $S_a$ and sixteen from $S_\da$. The geometry of this configuration
is ``degenerate'' in the sense that the impact parameter is necessarily zero
as we consider the scattering off an eight dimensional object in nine
dimensional space. The collision is therefore ``head on'' and one might be
skeptical about the validity of the eikonal approximation. We shall argue
however that, due to the cancellations involved, it is possible to
extrapolate the technique to this situation.

The background gauge $\bar A_9 = (i/2) vt\;\sigma^2$ represents a
D-particle moving towards the plane. This yields sixteen massive
bosonic degrees of freedom
($x_i$ and $\tilde x_i$) and sixteen fermionic ones ($S_\da$).
There are no massive
modes coming from gauge fields, ghosts or $S_a$ because the gauge group is
abelian. The contribution to the one loop integral\footnote{The ``would be''
phase shift if it were possible to set $b\not =0$.}
coming from the 8-orientifold alone
is thus ($\tau = i t$ and $\gamma = -iv$):
\beq
    \mathcal{A}_\mathrm{orientifold}=
    \mathrm{det}(-\partial_\tau^2 + \gamma^2\tau^2)^{-8}
    \mathrm{det}\pmatrix{
           \partial_\tau & \gamma\tau \cr
           \gamma\tau    & \partial_\tau \cr}^4. \label{orientifold}
\eeq
Expression (\ref{orientifold}) is badly divergent, but we can improve the
situation by considering the effective potential
$\mathcal{V}_\mathrm{orientifold}(v,r)$ defined through
\beq
    \mathcal{A}_\mathrm{orientifold}=
    \int_{-\infty}^{+\infty}dt\;
    \mathcal{V}_\mathrm{orientifold}(v,vt),
\eeq
that is, using analytic continuation for the integrals,
\beq
    \mathcal{V}_\mathrm{orientifold}(v,r) = v\int_0^{+\infty}\;
    \frac{ds}{\sqrt{\pi s}}e^{-sr^2}\frac{4 - 2\cos vs}
    {\sin vs} \approx r\left( -4 +
    \frac{v^2}{2r^4}+O\left(\frac{v^4}{r^8}\right) \right).
\eeq
However, this is not the end of the story. One must also take into
account the 16 fermionic modes from the D8-branes and their mirrors.
This gives a contribution
\beq
    \mathcal{A}_\mathrm{D8-brane}=
    \mathrm{det}\pmatrix{
           \partial_\tau & \gamma\tau \cr
           \gamma\tau    & \partial_\tau \cr}^4, \label{D8-brane}
\eeq
that always goes together with (\ref{orientifold}), yielding an extra
term in the effective potential:
\beq
    \mathcal{V}_\mathrm{D8-brane}(v,r) = -v\int_0^{+\infty}\;
    \frac{ds}{\sqrt{\pi s}}e^{-sr^2}\frac{2\cos vs}
    {\sin vs} \approx r\left( 4 +
    \frac{v^2}{2 r^4}+O\left(\frac{v^4}{r^8}\right) \right).
\eeq
Adding the two contributions together the velocity independent
piece linear in $r$ cancels and the total
effective potential becomes, to leading order,
\beq
     \mathcal{V}(v,r)=\mathcal{V}_\mathrm{orientifold}(v,r)+
     \mathcal{V}_\mathrm{D8-brane}(v,r) \approx \frac{v^2}{r^3}+\cdots.
\eeq
What we see here is a reflection in the low energy theory
of the necessity of having eight D8-branes together with the 8-orientifold.
Note also that if we allow the D8-branes to be positioned off the
8-orientifold, the linear potential will still cancel since the center of
mass for the D8-branes and their mirror images is always right on the
8-orientifold. That the cancellation persists even when the D8-branes are
far away from the 8-orientifold/D-particle system is of course due to
the potential being precisely linear. It is important that the linear
potential cancels as we move away from the system.
We would otherwise risk a bad IR-divergence.

Another example where care is needed is the case of a
D-particle in the vicinity of a
single D8-brane. Naively there would be a linear force on the D-particle
when we integrate out the fermionic open strings. However, this system would
suffer from an IR-divergence.  A way to fix this is to put another D8-brane
at infinity. Its only effect would be to cancel the linear force and we
thereby reproduce the result of \cite{L2}.

It is important to observe that the result for the potential above, i.e.
$v^2/r^3$, is only true for small $r$. At long distance it is not
protected against corrections from excited open strings, and we expect that
it will smoothly go over into $v^2/r^7$. Note that this is
really an interaction between a D-particle and its mirror image. Following
\cite{DKPS} one can deduce that for a 0-0 system $v^4/r^7$ is protected,
for a 0-4, $v^2/r^3$ is protected, while for 0-8, $r$ is protected.

Moving on to a less pathological situation, let us consider the scattering
of two D-particles with velocity $v$ and impact parameter $b$
in the 8 dimensional space parallel to the plane.
Let us therefore consider the background gauge $\bar X_1 = vt M^0$
and  $\bar X_2 = b M^0$ in the $SO(4)$ system of section 4 ($M^0$
given by (\ref{mzero})). From
(\ref{action}) we can read off the
massive modes and their contribution to the phase shift
\footnote{Actually, the most practical way to determine the masses of the
massive modes, is to make a group theoretical argument,
like the one in \cite{Par}, based on the weights of the representations.}:
\beqs
    &&e^{i\delta} = \mathrm{det}(-\partial_\tau^2 +
                \gamma^2\tau^2 + b^2 )^{-12}
    \mathrm{det}(-\partial_\tau^2 +
              \gamma^2\tau^2 + b^2 + 2\gamma)^{-2}
    \nonumber \\
    &&\times \mathrm{det}(-\partial_\tau^2 +
           \gamma^2\tau^2 + b^2 - 2\gamma)^{-2}
    \mathrm{det}\pmatrix{
           \partial_\tau & \gamma\tau -ib \cr
           \gamma\tau +ib   & \partial_\tau \cr}^{16}.\label{shift}
\eeqs
But $\delta$ from (\ref{shift}) is just twice the
phase shift computed in \cite{DKPS}! We
can therefore carry through all their results, in particular the flatness
of the metric on the moduli space. Actually, as we move away from the
8-orientifold, half of these modes become heavy and decouple, yielding
the type IIA result of \cite{DKPS}.

\section{The heterotic string}

In the introduction we promised that our results would also tell us
something about the heterotic string. Let us begin by a discussion of
what we might expect to find.

The system we are considering is, from M-theory point of view,
a compactification
from eleven to nine dimensions.  Note
that nine is the minimal dimension where we can find
all string theories through M-theory compactification.

There are two main types of such compactifications. The simplest is
to put M-theory on a torus, $S^1 \times S^1$, or equivalently type IIA on a
circle. T-duality then gives you the IIB string.
The D-particles that were studied in
\cite{us,KP,DKPS} clearly survives this compactification and can be
further studied in this nine dimensional setting. The other compactification
involves a $Z_2$ projection on one of the $S^1$ in the torus. A type IIB
string becomes a type I and a type IIA a type I'. The latter
is the system that we have been studying in this paper.
Exchanging the $S^1/Z_2$
and the $S^1$ produces the heterotic strings.

A membrane in M-theory can give rise to several different states in
the nine dimensional string theory with masses given by, in the notation
of \cite{HW2},
\beq
    M^2 = \frac{l^2}{R_{10}^2} +\frac{m^2}
   {R_{11}^2}+ n^2 R_{10}^2 R_{11}^2   \label{tjo1}
\eeq
in ``M-theory'' units.
This is true both for the type II side and the heterotic/type I side.
In the latter case we must make a $Z_2$ projection on the tenth
direction.
Depending on the specific string theory the states have very
different interpretations. In I' units the masses are
\beq
    M^2_{I'} = \frac{l^2}{R_{I'}^2} +
    \frac{m^2}{\lambda _{I'}^2}+ n^2 R_{I'}^2, \label{tjo2}
\eeq
while in heterotic $E_8 \times E_8$ we have
\beq
    M^2_{E_8} = \frac{l^2}{\lhet^2}
    +\frac{m^2}{\rhet^2}+ n^2 \rhet^2.
\eeq
In type I' we have two 8-orientifolds separated by $R_{I'}$, one for each
fixed point of $Z_2$. In addition there
are 16 D8-branes whose positions determine the space time gauge group. The
coupling constant $\lop$
is determined by the remaining compact $S^1$. In the case
of the heterotic string, the coupling constant $\lhet$
is instead determined by
the distance between the two 8-orientifolds.
Our D-particle, which has a non zero $m$ quantum number,
is hence a nonperturbative state in the I' theory but just
a Kaluza-Klein mode of the heterotic string in the heterotic theory.

The D-particles will interact with each other, with their mirror images,
and also with the the D8-branes and their mirror images.
States with $l$ quantum numbers are not stable in the heterotic/I case.
This is due to momentum not being conserved in the tenth direction; here we
have an interval not a circle. This is reflected in e.g. a velocity
dependent force transverse to the 8-orientifolds. This we indeed verified
in the previous section.

The calculations that we performed in the previous section for scattering
parallel to the plane is effectively nine dimensional, and if we can extend
them to include two planes we should be able to interpret the results
in terms of a nine dimensional heterotic string theory. What do we know
about the dynamics of black holes in such a theory?

Let us consider the (gravitational)
bosonic part of the effective action for a heterotic
string in nine dimensions. This is obtained through dimensional reduction
and is given by
\beq
   S= \frac{1}{2} \int d^9 x \sqrt{-g}
   \left( R - \frac{1}{2} (d \xi )^2 -
   \frac{1}{2} (d \phi )^2 -
   \frac{1}{4}e^{-\frac{4}{\sqrt{7}}\xi} (d {\cal A})^2 -
   \frac{1}{4} e^{-\phi + \frac{3}{\sqrt{7}} \xi} (dA)^2 \right)
\eeq
where $\xi\sim g_{99}$,
${\cal A} _\mu \sim g_{\mu 9}$ and $A_\mu \sim B_{\mu 9}$ and we have set
the two form $B_{\mu\nu}=0$ since we are interested in $p=0$ solitons.
One-scalar-solutions can be obtained in two different ways, either by taking
a linear combination of $\phi$ and $\xi$ or by just setting $\phi=A=0$.
The first of these has $\Delta =2$, the second $\Delta =4$, where
$\Delta$ is an invariant given by \cite{DKL, LPSS, LP}
\beq
    \Delta = a^2 +\frac{2(p+1)(D-p-3)}{D-2} = a^2 + \frac{12}{7},
\eeq
for $p=0$ and $D=9$.
The quantity $a^2$ is obtained from
$e^{-\mathbf{a}\cdot\mathbf{\Phi}}$, where $\mathbf{\Phi}
= (\phi , \xi )$. The linear combination $\mathbf{a} \cdot\mathbf{\Phi}$
is the scalar field that we allow to be nonzero. One also has that
$\Delta = 4/N$ where $N$ is the number of participating field strengths.
The black hole with $\Delta=4$ is the one which is described by a
D-particle because it's a purely Kaluza-Klein mode.

What will the force between two such black holes be?
We will follow \cite{DKL}
and find that the force starts only at $v^4$. This is a general result
for $\Delta=4$ as can be seen easily from the following calculation.
The metric of an extremal black hole in $D$ dimensions is, (see e.g.
\cite{HS1,KT2}),
\beq
   ds^2 = f^{\frac{N}{D-2}} (-f^{-N} dt^2
   +dr^2 +r^2 d \Omega ^2 _{D-2})   ,
\eeq
where
\beq
   f= 1+ \frac{r_e^{D-3}}{r^{D-3}}   .
\eeq
The dilaton is given by
\beq
    e^{2\phi} = f^{-aN} .
\eeq
The action for a black hole moving in the background of another,
identical black hole, is
\beq
    S= -m\int dt \left(\sqrt{-g_{\mu \nu} \dot{x}^{\mu}
    \dot{x}^{\nu} e^{a\phi}}-\dot{x}^{\mu} A_{\mu}\right).
    \label{sig}
\eeq
{}From (\ref{sig}) we can read off the force. By supersymmetry the gauge field
$A_{\mu}$ is such that the velocity independent force cancels.
Let us concentrate on velocity dependent forces. We find:
\beq
   \sqrt{-g_{\mu \nu} \dot{x}^{\mu} \dot{x}^{\nu} e^{a\phi}} \sim
   g_{tt}^{1/2} e^{a\phi /2} - \frac{1}{2} e^{a\phi /2}
   \frac{g_{rr}}{g_{tt}^{1/2}} v^2 + O(v^4 ).
\eeq
The coefficient of $v^2$ is hence given by
\beq
    -\frac{1}{2} f^{\frac{N}{4} (4-\Delta )}  ,
\eeq
and thereby we find that the force vanishes for $\Delta =4$ but is
nonzero for $\Delta \neq 4$.

This is consistent with what we calculated in the previous section
for scattering parallel to the 8-orientifold.
In the heterotic limit there is an extra subtlety due to the fact that we have
two 8-orientifolds that  come close.
This makes it possible to have extra open strings
stretching between the D-particles.
We must in fact sum over strings
that bounce between the two 8-orientifolds an arbitrary number of times. With
one 8-orientifold we only needed to consider a string stretching directly
between
the two D-particles and one that bounced off
the 8-orientifold once (equivalently connected to the
mirror image of the other D-particle). This sum converts the $1/r^7$ to
$1/r^6$ at weak coupling.

Let us recall the table in the previous section!
The fact that the length of the interval $R_{10}$ does not appear in the last
column, makes it more plausible to extrapolate the results to the
heterotic case. The fact that
the orbital momentum is of order one in M-theory units is the direct
analogue of the fact noted in \cite{KP, DKPS} that the D-particles allow
one to probe distances of the
order of the eleven dimensional Planck length, which is smaller than the
string length in \emph{both} the type I' and the heterotic limit:
$l_{\mathrm{min}} = \lop^{1/3} << 1$
in the limit where the type I' description applies and
$l_{\mathrm{min}} = \lhet^{1/3} << 1$ in the limit where the heterotic
string is weakly coupled.

One should note that the calculation is valid only for $v$ small
(we are considering Yang-Mills, not Born-Infeld), which for the
heterotic string means that $\rhet/\lhet^{1/3}$ is small.
Luckily, in nine dimensions, there exists a region where
both the type I' and the heterotic string
are weakly coupled. But since $\rhet = \lop ^{2/3} l_{11}$ and
$R_{I'} = \lhet ^{2/3} l_{11}$ it follows that the compact dimension must
be smaller than the eleven dimensional
Planck length for a weakly coupled type I', and the same must be true for the
distance between the orientifolds to also have a weakly coupled
heterotic string. While our quantum mechanical system still should provide
a good description, the Born-Oppenheimer approximation would break down
for strings stretching from a D-particle to its own mirror. However, these
strings were only relevant for transverse scattering and not for the parallel
scattering that we are considering here, so that there is some hope
that some of these results carry over to the heterotic string.

The black hole with $\Delta =2$ is perhaps even more interesting since it
has a non flat moduli space. This black hole must correspond
to a heterotic string that has both winding and momentum in the compact
direction since it couples to both $A$ and ${\cal A}$.
Such an example was constructed in \cite{GHRW,DGHW} where, indeed, a
non flat moduli space was verified. If we consider the mass
formulae above we see that such a system must correspond to a membrane
wrapped
around $R_{11}$ and stretched between the 8-orientifolds.
Can such an object be constructed using
D-particles?

In \cite{us} a bound state of two D-particles was studied and it was found
that the bound state had a total mass given by
\beq
    \frac{2}{R_{11}} + \epsilon R_{11}
\eeq
in M-theory units. The second term is due to the
presence of a string stretching between the two D-particles.
Here we would like, in analogy with (\ref{tjo1}) and (\ref{tjo2}),
to interpret the string as
a membrane wound around the compact dimension. $\epsilon$ above is the
expectation value of the distance between the D-particles and
$\epsilon R_{11}$ is hence the area of the membrane. We conclude, then,
that a string stretching between two D-particles is actually a membrane
wrapped around $R_{11}$. Note that this
state is a non-BPS state, with no topological protection and therefore not
stable. It is interesting to note that what is essentially only a
scaling argument, provides evidence that the D-particle system not only
knows about M-theory through the eleven dimensional Planck length, but also
about membranes.

With these observations in mind, it is reasonable to
expect that it should be possible to construct stable states with
both winding and momentum
corresponding to the $\Delta =2$ black holes using D-technology.
We should then have strings stretching
between the 8-orientifolds, threaded by D-particle beads.

\section{Conclusions}

In this paper we have investigated the quantum mechanical system that
describes D-particles in the presence of 8-orientifolds and D8-branes.
For consistency each 8-orientifold must be matched by 8 D8-branes to
cancel any linear potential. This we saw reflected in the
quantum mechanics of the D-particles.

We have calculated the scattering of D-particles moving parallel to the
8-orientifold/D8-brane system finding $v^4/r^7$ forces only. In addition
we considered D-particles moving in the transverse direction. In that case
the force turned out to be proportional to $v^2/r^3$.

While this is obviously an example of type I' dynamics, we have argued that
the result also can be interpreted as a calculation of the scattering of
$\Delta =4$, heterotic Kaluza-Klein black holes.

\section*{Acknowledgements.}

We would like to thank P. Stjernberg for valuable discussions and
G. Lifschytz for e-mail communications.


\begin{thebibliography}{99}

\bibitem{P1} J. Polchinski, Phys. Rev. Lett. {\bf 75} (1995) 4724,
{\tt hep-th/9510017}.

\bibitem{PCJ} J. Polchinski, S. Chaudhuri and C.V. Johnson,
``Notes on D-branes'', \hfill\break {\tt hep-th/9602052}.

\bibitem{HT} C.M. Hull and P.K. Townsend, Nucl. Phys. {\bf B438} (1995)
109, \hfill\break {\tt hep-th/9410167}.

\bibitem{T} P.K. Townsend, Phys. Lett. {\bf B350} (1995) 184, {\tt
hep-th/9501068}.

\bibitem{W1} E. Witten, Nucl. Phys. {\bf B443} (1995) 85, {\tt
hep-th/9503124}.

\bibitem{PW} J. Polchinski and E. Witten, Nucl. Phys. {\bf B460} (1996) 525,
\hfill\break {\tt hep-th/9510169}.

\bibitem{S6}  J.H. Schwarz, Phys. Lett. {\bf B367} (1996) 97,
{\tt hep-th/9510086}.

\bibitem{HW2} P. Horava and E. Witten, Nucl. Phys. {\bf B460} (1996) 506,
\hfill\break {\tt hep-th/9510209}.

\bibitem{W2} E. Witten, Nucl. Phys. {\bf B460} (1996) 541,
{\tt hep-th/9511030}.

\bibitem{T2} P.K. Townsend, Phys. Lett. {\bf B373} (1996) 68,
{\tt hep-th/9512062}.

\bibitem{HW1} P. Horava and E. Witten, Nucl. Phys. {\bf B475} (1996) 94,
{\tt hep-th/9603142}.

\bibitem{DKL}  M.J. Duff, R.R. Khuri and J.X. Lu,
Phys. Rep. {\bf 259} (1995) 213, \hfill\break {\tt hep-th/9412184}.

\bibitem{P2} J. Polchinski, ``String Duality: a Colloquium'',
{\tt hep-th/9607050}.

\bibitem{S7} J.H. Schwarz, ``Lectures on Superstring and M-theory Dualities'',
{\tt hep-th/9607201}.

\bibitem{D2} M.J. Duff, ``M-theory (The Theory Formerly Known as Strings)'',
{\tt hep-th/9608117}.

\bibitem{D3} M.R. Douglas, ``Superstring Dualities, Dirichlet Branes and the
Small Scale Structure of Space'', {\tt hep-th/9610041}.

\bibitem{DLP} J. Dai, R.G.  Leigh and J. Polchinski,
Mod. Phys. Lett. {\bf A4} (1989) 2073.

\bibitem{L1} R. G. Leigh, Mod. Phys. Lett. {\bf A4} (1989) 2767.

\bibitem{G} M.B. Green, Phys. Lett. {\bf B266} (1991) 325.

\bibitem{SV} A. Strominger and C. Vafa, Phys. Lett. {\bf B379} (1996) 99,
\hfill\break {\tt hep-th/9601029}.

\bibitem{DM} S.R. Das and S.D. Mathur, ``Excitations of D-strings, Entropy
and Duality'', {\tt hep-th/9601152}.

\bibitem{CM} C.G. Callan and J.M. Maldacena, Nucl. Phys. {\bf B472} (1996) 591,
{\tt hep-th/9602043}.

\bibitem{HS2} G.T. Horowitz and A. Strominger, Phys. Rev. Lett. {\bf 77} (1996)
2368, {\tt hep-th/9602051}.

\bibitem{GKP} S.S. Gubser, I.R. Klebanov and A.W. Peet, Phys. Rev. {\bf D54}
(1996) 3915, {\tt hep-th/9602135}.

\bibitem{M} S.D. Mathur, ``Non BPS Excitations of D-branes and Black Holes'',
{\tt hep-th/9609053}.

\bibitem{KT} I.R. Klebanov and L. Thorlacius, Phys. Lett.
{\bf B371} (1996) 51, \hfill\break {\tt hep-th/9510200}.

\bibitem{GHKM} S.S. Gubser, A. Hashimoto, I.R. Klebanov and J.M. Maldacena,
Nucl. Phys. {\bf B472} (1996) 231, {\tt hep-th/9601057}.

\bibitem{HK} A. Hashimoto and I.R. Klebanov, Phys. Lett. {\bf B381} (1996) 437,
{\tt hep-th/9604065}.

\bibitem{H2} A. Hashimoto, ``Dynamics of Dirichlet-Neumann Open Strings on
D-branes'',  {\tt hep-th/9608127}.

\bibitem{us}  U.H. Danielsson, G. Ferretti and B. Sundborg, ``D-particle
Dynamics and Bound States'', Int. J. Mod. Phys. (to appear),
{\tt hep-th/9603081}.

\bibitem{KP} D. Kabat and P. Pouliot, Phys. Rev. Lett. {\bf 77} (1996) 1004,
\hfill\break {\tt hep-th/9603127}.

\bibitem{DL} M.R. Douglas and M. Li, ``D-brane Realization of $N=2$ Super
Yang-Mills Theory in Four Dimensions'', {\tt hep-th/9604041}.

\bibitem{D1} M.R. Douglas, ``Gauge Fields and D-branes'', {\tt hep-th/9604198}.

\bibitem{BDS} T. Banks, M.R. Douglas and N. Seiberg, ``Probing F-theory with
Branes'', {\tt hep-th/9605199}.

\bibitem{DKPS} M.R. Douglas, D. Kabat, P. Pouliot
and S.H. Shenker, ``D-branes and
Short Distances in String Theory'', {\tt hep-th/9608024}.

\bibitem{S1} N. Seiberg, ``IR Dynamics on Branes and Space-time Geometry'',
{\tt hep-th/9606017}.

\bibitem{BFSS} T. Banks, W. Fischler,
S.H. Shenker and L. Susskind,  ``M-theory as
a Matrix Model: A Conjecture'', {\tt hep-th/9610043}.

\bibitem{H1} C.M.  Hull, Nucl. Phys. {\bf B468} (1996) 113,
{\tt hep-th/9512181}.

\bibitem{S3} A. Sagnotti, in Cargese '87, ``Non-perturbative Quantum
Field Theory'' ed. G. Mack et. al. (Pergamon Press, 1988).

\bibitem{H3} P. Horava, Nucl. Phys. {\bf B327} (1989) 461.

\bibitem{PS}  G. Pradisi and A. Sagnotti, Phys. Lett. {\bf B216} (1989) 59.

\bibitem{GP} E.G.Gimon and J. Polchinski, Phys. Rev. {\bf D54} (1996)
\hfill\break 1667, {\tt hep-th/9601038}.

\bibitem{W3} E. Witten, Nucl. Phys. {\bf B460} (1996) 335,
{\tt hep-th/9510135}.

\bibitem{BSS} L. Brink, J.H. Schwarz and J. Scherk,
Nucl. Phys. {\bf B121} (1977) 77.

\bibitem{S4}  A. Sen, Phys. Rev. {\bf D54} (1996) 2964, {\tt hep-th/9510229}.

\bibitem{S5} A. Sen, Mod. Phys. Lett. {\bf A11} (1996) 827,
{\tt hep-th/9512203}.

\bibitem{B} C. Bachas, Phys. Lett. {\bf B374} (1996) 37, {\tt hep-th/9511043}.

\bibitem{BS} T. Banks and L. Susskind, ``Brane-Antibrane Forces'',
{\tt hep-th/9511194}.

\bibitem{FPR} W. Fischler, S. Paban and M. Rozali, Phys. Lett. {\bf B381}
(1996) 62, {\tt hep-th/9604014}.

\bibitem{L2} G. Lifschytz, ``Comparing D-branes to Black-branes'',
{\tt hep-th/9604156}.

\bibitem{BCFPR}  D. Berenstein, R. Corrado, W. Fischler, S. Paban and
M. Rozali, ``Virtual D-branes'', {\tt hep-th/9605168}.

\bibitem{S2} S.H. Shenker, ``Another Length Scale in String Theory?'',
\hfill\break {\tt hep-th/9509132}.

\bibitem{Par} U.H. Danielsson and P. Stjernberg, Phys. Lett. {\bf B380}

(1996) 68, {\tt hep-th/9603082}.

\bibitem{LPSS} H. Lu, C.N. Pope, E. Sezgin and K.S. Stelle, Nucl. Phys.
{\bf B456} (1995) 669, {\tt hep-th/9508042}.

\bibitem{LP} H. Lu and C.N. Pope, Nucl. Phys. {\bf B465} (1996) 127,
{\tt hep-th/9512012}.

\bibitem{HS1} G.T. Horowitz and A. Strominger, Nucl. Phys. {\bf B360} (1991)
197.

\bibitem{KT2} I.R. Klebanov and A.A. Tseytlin, Nucl. Phys. {\bf B475} (1996)
164, {\tt hep-th/9604089}.

\bibitem{GHRW} J.P. Gauntlett, J.A. Harvey, M.M. Robinson and D. Waldram, Nucl.
Phys. {\bf B411} (1994) 461, {\tt hep-th/9305066}.

\bibitem{DGHW} A. Dabholkar, J.P. Gauntlett, J.A. Harvey and D. Waldram,
Nucl. Phys. {\bf B474} (1996) 85, {\tt hep-th/9511053}.

\end{thebibliography}
\end{document}